\documentclass[aps,prl,showpacs,twocolumn]{revtex4}
\usepackage{graphicx}
\usepackage[T1]{fontenc} 
\usepackage{epsfig,latexsym}
\usepackage{color}
\usepackage[latin1]{inputenc}
\usepackage{amsmath}
\usepackage{amssymb}
\usepackage{amsfonts}
\usepackage{indentfirst}
\usepackage{hyperref}
\usepackage{MnSymbol,wasysym}
\usepackage{ae}
\usepackage{float}

\begin{document}
\title{Fractional Angular Momenta, Gouy and Berry phases in Relativistic Bateman-Hillion-Gaussian Beams of Electrons }
\author{Robert Ducharme$^{1}$}
\email{robertjducharme66@gmail.com}
\author{Irismar G. da Paz$^{2}$} 
\email{irismarpaz@ufpi.edu.br}
\author{Armen G. Hayrapetyan$^{3,4,5}$}
\email{armen.hayr@gmail.com}

\affiliation{$^{1}$ 2112 Oakmeadow Pl., Bedford, TX 76021, USA}

\affiliation{$^2$ Departamento de F\'{\i}sica, Universidade Federal
do Piau\'{\i}, Campus Ministro Petr\^{o}nio Portela, CEP 64049-550,
Teresina, PI, Brazil}

\affiliation{$^{3}$ d-fine GmbH, Bavariafilmplatz 8, 82031 Gr\"unwald, Germany}

\affiliation{$^{4}$ Mathematical Institute, University of Oxford,
Radcliffe Observatory Quarter, Woodstock Rd, Oxford OX2 6GG, UK}

\affiliation{$^{5}$  Max-Planck-Institut f\"ur Physik komplexer Systeme, N\"othnitzer Str. 38, 01187 Dresden, Germany}

\begin{abstract}
A new Bateman-Hillion solution to the Dirac equation for a relativistic Gaussian electron beam taking explicit account of the $4$-position of the beam waist is presented. This solution has a pure Gaussian form in the paraxial limit but beyond it contains higher order Laguerre-Gaussian components attributable to the tighter focusing. One implication of the mixed mode nature of strongly diffracting beams is that the expectation values for spin and orbital angular momenta are fractional and are interrelated to each other by \textit{intrinsic spin-orbit coupling}. Our results for these properties align with earlier work on Bessel beams [Bliokh \textit{et al.} Phys. Rev. Lett. \textbf{107}, 174802 (2011)] and show that fractional angular momenta can be expressed by means of a Berry phase. The most significant difference arises, though, due to the fact that Laguerre-Gaussian beams naturally contain Gouy phase, while Bessel beams do not. 
We show that Gouy phase is also related to Berry phase and that Gouy phase fronts that are flat in the paraxial limit become curved beyond it.

\end{abstract}

\pacs{41.85.-p, 03.65.Pm, 03.65.Vf,42.50.Tx}

\maketitle

{\textit{Introduction}.---} In the past decade, there has been considerable progress 
towards solving the Dirac equation (DE) for the purpose of unveiling detailed properties 
of electron vortex beams carrying both spin and orbital angular momenta. The earliest of this work has modelled Bessel beams  
as a linear superposition of 
Dirac or Dirac-Volkov~\cite{Bliokh:11, Hayrapetyan-Karlovets} plane waves
in contrast to non-relativistic Laguerre-Gaussian (LG)~\cite{Bliokh:07}
and Bessel beams
\cite{Schattschneider:11} acting as solutions to paraxial and non-paraxial wave equations, respectively.
More recently, the attention has been focused on
investigating symmetry properties of relativistic electrons  
to better understand the nature of their vortex formation \cite{Bliokh:17,Karlovets:18}
and to construct other types of wave packets~\cite{Karlovets:18,BB}
(see also the debate in Refs.~\cite{BB,SMB,BB-comment}).

A 3-vector position in space is not form invariant under Lorentz transformations. For fully relativistic calculations it is therefore necessary to replace points using point events (4-positions) so that a beam front with velocity u that reaches the waist at time T will have time $t = T + \xi_B / u$ for any other point at a distance $\xi_B$ before $(\xi_B<0)$ or after $(\xi_B>0)$ the waist. The use of point events as energy-momentum waypoint markers has precedent in the derivation of Lienard-Wiechert potentials \cite{SR} and constraint mechanics \cite{AK,CA}. Beam solutions that incorporate a 4-position beam waist have been shown to reduce to traditional beam models \cite{RD1, RD2, RD3} in the non-relativistic limit. Beam solutions that zero out the beam waist as the origin of the beam coordinate system thus carry a hidden non-relativistic 3-vector.

In a typical electron microscope assembly, 
a Gaussian beam passes from an electron gun 
to a magnetic lens that focuses it to a small waist diameter. Assuming a modest current 
of energetic electrons ($\sim$ 100keV), the average separation between them will be large 
enough so that electron repulsion can be ignored. Under these conditions, the expected 
diameter of the beam waist will be about a hundred times the wavelength of the electrons 
unless corrective measures are taken to reduce the strong spherical and chromatic
aberration, which is a normal feature of magnetic lenses \cite{RE}.
Such transmission electron microscopes are by far the only tool
to produce electron vortex beams~\cite{EVB-exp}, which are nowadays widely used in various physical setups~\cite{review-EVB}. This includes vortices in external fields~\cite{EVB-in-external-fields},
scattering~\cite{EVB-with-scattering}, atomic 
processes~\cite{EVB-with-radiation} and indicates a further possible application for electron beams to trap 
(sub)nanoparticles in a close analogy 
with optical trapping via light vortices~\cite{manipulation-via-twist}.

Over the course of last two-three decades, advances in optical instrumentation have made it possible to produce light vortex beams, which typically carry an integer number of orbital angular momentum (OAM) quanta 
\cite{review-twisted-light,twisted-light-other}.
This led to 
an elucidation of the nature of \textit{fractional} OAM (fOAM) including its clear understanding in terms of Berry phase \cite{FOAM,FOAM-exp}.

Our intention here is to build on existing work both on optical and electron vortices to calculate fractional -- spin and orbital -- angular momenta  
for a tightly focused relativistic Gaussian beam of electrons, which contains higher order LG modes. Taking proper account of the 4-position of the beam waist, we evaluate the total energy-momentum of relativistic electrons and derive a fundamental property called the Gouy phase
\cite{gouy1,gouy2}, accumulated along the propagation direction as a result of the transverse localization of the beam~\cite{feng2001-yang2006}. We demonstrate the interplay of the Gouy phase and the fractional angular momenta, quantified by an intrinsic spin-orbit interaction (SOI) term,
with the Berry geometric phase. In view of this, we parameterize the total \textit{shift} of the Gouy phase in a relativistic Gaussian beam, from far field to far field, in terms of the Berry phase. We shall also calculate beyond the paraxial limit to show that Gouy phase fronts that have generally thought to be planar are actually curved.

{\textit{Exact Bateman-Hillion-Gaussian beams from Dirac equation}.---}
In order to achieve our goal, we develop a theory that incorporates Lorentz invariance of relativistic Gaussian solutions of wave equations known 
as Bateman-Hillion (BH) solutions \citep{BA-PH,BS-APK} that also take proper account of the beam waist position. We use the BH ansatz to solve the Klein-Gordon equation (KGE) then convert it to a solution of the full DE. Our solutions build on existing approaches for relativistic LG beams~\cite{Karlovets:18,BB}. It must be recognized, however, that our inclusion of the beam waist 4-position matters since it brings a resolution to a well known problem of BH Gaussian beams that electrons move in both directions \cite{BB} additional to certain other benefits that include Lorentz invariance and correspondence to accepted beam models in the non-relativistic and paraxial limits.

Consider a beam of electrons each having a mass of $m$, a $4$-position
$x_\mu = \left( ct, - \mathbf{r} \right)$ and a $4$-momentum
$p_\mu = \left( E / c , - \mathbf{p} \right)$, where $\mu = \{0,1,2,3 \}$
and $c$ is the speed of light in vacuum. It follows the particle has an
energy $E$ and 3-momentum $\mathbf{p}$ at world time $t$ and world position
$\mathbf{r}$. Let us also assume that each electron passes through a beam
waist with a $4$-position $X_\mu = \left(cT, - \mathbf{R} \right)$, where
$\mathbf{R}$ is the world position of the beam waist at world time $T$.
Note that we introduce two different time coordinates since the equality $T=t$
is not form preserving under Lorentz transformations. 

The dynamics of each of the electrons in the beam is then
given by the DE expressed as~\cite{PAMD}
\begin{equation} \label{eq: dirac_eq}
(\gamma^\mu \hat{p}_\mu - m c) \, \Psi_{\pm} (x_\mu, X_\mu) \,\, = \,\, 0 \, .
\end{equation}
Here,
$\hat{p}_\mu = \imath \hbar \partial / \partial x^\mu$ is
the canonical $4$-momentum operator, $\gamma^\mu$ are the
Dirac matrices, $\hbar$ is the reduced Planck's constant,
while $\Psi_{\pm}(x_\mu, X_\mu)$ represents a bi-spinor wave function
of each individual electron, where ``$\pm$'' stand for 
spin-up and -down states, respectively. Equation (\ref{eq: dirac_eq})
also has two negative-energy bi-spinor solutions that we will not
consider since they describe anti-particles.
The DE~(\ref{eq: dirac_eq}) can be simplified using the substitution
\begin{eqnarray} \label{eq: bispinor}
\Psi_\pm \,  (x_\mu, X_\mu) \,\, = \,\, \left[ \begin{array}{c}
(\hat{p}_0+m c) \chi_{\pm} \\
\sigma_i \hat{p}_i \chi_{\pm}
\end{array} \right]
\Psi \,  (x_\mu, X_\mu)
\end{eqnarray}
with $\chi_+ = \left( 1 \,\, 0 \right)^T$, $\chi_- = \left( 0 \,\, 1 \right)^T$
being two-component spinors, $\Psi (x_\mu, X_\mu)$ a scalar function, $\sigma_i$ the Pauli matrices
($i = \{ 1 , 2 , 3 \}$), and where the inner product 
$\sigma_i \hat{p}_i \equiv \sigma_1 \hat{p}_1 + \sigma_2 \hat{p}_2 + \sigma_3 \hat{p}_3$ and the superscript T means ``transposed''.
Combining Eqs.~(\ref{eq: dirac_eq}) and (\ref{eq: bispinor}) leads
to the KGE for $\Psi$
\begin{equation} \label{eq: KG}
(\hat{p}_\mu \hat{p}^\mu - m^2c^2)\Psi \, (x_\mu, X_\mu) \,\, = \,\, 0 \, .
\end{equation}
The clear understanding here is that the bi-spinor solution $\Psi_{\pm}$ satisfies
the DE provided that the scalar function $\Psi$ acts as a solution of the KGE, a procedure that is also applied in Ref.~\cite{BB} for constructing relativistic wave packets with non-zero OAM.

The solution to the KGE~(\ref{eq: KG}) for the Gaussian beam has been developed
in two recent papers \cite{RD1, RD2}. For our purposes, we start from the BH
based ansatz
\begin{equation} \label{eq: bateman_ansatz}
\Psi \, (x_\mu, X_\mu) \,\, = \,\, C \Phi(\xi_1,\xi_2, \xi_3+\xi_0)\exp \left( -\imath k_\mu^\prime x^\mu \right),
\end{equation}
where $C$ is a constant number, $\xi_\mu = x_\mu - X_\mu$ is the $4$-position of the electron
relative to the beam waist, $k_\mu^\prime$ is the wave $4$-vector and $\Phi(\xi_1,\xi_2, \xi_3+\xi_0)$ is a scalar function incorporating non-trivial vortex and both space- and time-dependent phase structures of the electron beam.

Following Ref \cite{RD2}, we insert the BH ansatz
(\ref{eq: bateman_ansatz}) into the KGE~(\ref{eq: KG}) and
solve the resulting equation for $\Phi( \xi_\rho, \xi_\phi, \xi_3 + \xi_0)$ by utilizing the `radial' $\xi_\rho = \sqrt{\xi_1^2+\xi_2^2}$ and `azimuthal'
$\xi_\phi = \arctan (\xi_2 / \xi_1)$ coordinates. This leads to the LG solution
\begin{eqnarray}
\label{eq: laguerre_gauss_solution}
\Phi_{lp} =  a_{\ell p}\left(
\frac{\sqrt{2}\xi_\rho}{|w|}\right)^{|l|}
\!\!\! L_p^{|l|}\left(
\frac{2 \xi_\rho^2}{|w|^2}\right)
\exp \left( \! -\frac{ \xi_\rho^2}{w_0 w} +\imath l \xi_{\phi}- \imath
g_{lp} \right) \,
\end{eqnarray}
with $a_{\ell p} \equiv \sqrt{ 2p! / \left[\pi |w|^2(p+|l|)!\right]}$ and
$k_\mu^\prime k^{\mu \prime} = m^2 c^2/\hbar^2$.
Furthermore, $L_p^{|l|}$ represent the generalized Laguerre polynomials in terms of the radial, $p \geq 0$, and the azimuthal indices, $-\infty < l < \infty$,
\begin{equation} \label{eq: gouy_phase_lp}
g_{lp} \,\, = \,\, (1+|l|+2p)\arctan [ 2 \kappa (\xi_3+\xi_0) ]
\end{equation}
is the Gouy phase and $\kappa = [w_0^2(k_3+k_0)]^{-1}$.
The solution (\ref{eq: laguerre_gauss_solution}) also contains the complex parameter
\begin{equation} \label{eq: complex_beam_parameter}
w \,\, = \,\, w_0\left[1+ 2\kappa (\xi_3+\xi_0)\imath \right] \, ,
\end{equation}
whose modulus, $|w|$, characterizes the beam radius, such that $w_0$ represents the beam radius at the waist. Note that the Gouy phase (\ref{eq: gouy_phase_lp}) depends on both the \textit{space and time} variables in sharp contrast to only a time-dependent Gouy phase of Ref. \cite{Karlovets:18}. Our setup considers a beam confined in the transverse $(x, y)$-plane and propagating in the longitudinal direction, chosen to be the $z$-axis, meaning that the beam spreads as a function of $z$ and time~$t$. This confinement in two dimensions can be experimentally designed by spherical lenses complementing the one-dimensional case realized by cylindrical lenses. The difference of expressions in round parentheses ($1$ in Eq. (\ref{eq: gouy_phase_lp}) and $3/2$ of Ref. \cite{Karlovets:18}) is a consequence of an altered scenario, when the beam is confined in three dimensions \cite{Karlovets:18}, a situation which is yet to be experimentally generated.

Equations (\ref{eq: bispinor}), (\ref{eq: bateman_ansatz}) and (\ref{eq: laguerre_gauss_solution}) constitute an
exact BH solution to the DE for LG modes of the electron beam. We shall, however, focus our attention only on the Gaussian wave packets leaving a treatment of
more general LG modes for future work as relevant physical conclusions can be drawn already from \textit{relativistic Bateman-Hillion-Gaussian (BHG) beams}, i.e., when $\ell = p = 0$.
Inserting, therefore, $\Psi =\Psi_{00} = C \Phi_{00}\exp \left( -\imath
k_\mu^\prime x^\mu \right)$ into Eq.~(\ref{eq: bispinor}) leads to
\begin{equation} \label{eq: bispinor_explicit}
\Psi_\pm = \left( \begin{array}{c}
b\chi_{\pm} \\
\pm \hbar k_3 \chi_{\pm}
\end{array} \right) \Psi_{00}
+ \left( \begin{array}{c}
\hbar \kappa \chi_{\pm}  \\
\pm \hbar \kappa \chi_{\pm}
\end{array} \right) \Psi_{01}
\pm \left( \begin{array}{c}
0 \\
\frac{\sqrt{2} \hbar}{w_0} \chi_{\mp}
\end{array} \right) \Psi_{ 10} \, ,
\end{equation}
where $k_\mu = (k_0^\prime +\kappa, 0, 0,-k_3^\prime +\kappa)$ will be referred to as the effective wave vector of the electron beam,
\begin{eqnarray}
\nonumber
\Psi_{00} & = &
	\frac{\sqrt{2} C}{\sqrt{\pi} w}
	\exp \left( - \frac{\xi_\rho^2}{w_0 w} \right)
	\exp \left( - \imath k_\mu^\prime  x^\mu \right) \, ,
\\
\nonumber
\Psi_{01} & = &
	\frac{\sqrt{2} C}{\sqrt{\pi} w}
	\left( \frac{\left|w\right|^2}{w^2} - \frac{2 \xi_\rho^2}{w^2} \right)
	\exp \left( - \frac{\xi_\rho^2}{w_0 w} \right)
	\exp \left( - \imath k_\mu^\prime  x^\mu \right) \, ,
\\
\nonumber
\Psi_{10} & = &
	\frac{2 C \xi_\rho}{\sqrt{\pi} w^2}
	\exp \left( - \frac{\xi_\rho^2}{w_0 w} + \imath \xi_\phi \right)
	\exp \left( - \imath k_\mu^\prime  x^\mu \right) \, ,
\end{eqnarray}
$b \equiv \hbar k_0 + mc$. We have omitted the arguments $(x_\mu, \xi_\mu)$ for brevity. Equation (\ref{eq: bispinor_explicit}) is the exact solution to the DE
for the lowest order (Gaussian) bi-spinor mode of electron beam.
In the paraxial and semi-relativistic limit 
($k_3 \ll k_0$, $k_0 \simeq m c$), we recover Barnett's 
solution~\cite{SMB}.

The constant $C$ in Eq.(\ref{eq: bateman_ansatz}) can be determined
from the Dirac current $j_\mu^\pm = \Psi_{\pm}^{\dagger} \gamma_0 \gamma_\mu
\Psi_{\pm}$ using the normalization condition,
$ \langle j_\mu^\pm \left(\xi_1 , \xi_2 , \xi_3 + \xi_0 \right) \rangle = k_\mu/k_0 $~\cite{footnote1},
which implies $C = \left[2(\hbar k_0 b +\hbar^2 \kappa^2)\right]^{-1/2} $
and gives the expected velocity of the beam front $\xi_B/\xi_0 = k_3/k_0 $ where $\xi_B = \xi_3 \sqrt{1 + \xi_\rho^2 / \xi_3^2}$ is distance traveled.
We now eliminate the elapsed time $\xi_0$ in Eq. (\ref{eq: gouy_phase_lp}) using the previous expression to give
\begin{eqnarray} 
\label{eq: gouy_phase}
g_{lp} & = & (1+|l|+2p)\arctan \left[ \frac{k_3 \xi_3 +k_0 \xi_B}{\xi_R (k_3+k_0)} \right],
\end{eqnarray}
for $\xi_B >> \xi_R$ where $\xi_R = \frac{1}{2}k_3 w_0^2$ represents the Rayleigh range. To recover the standard LG beam formulae we shall instead set $\xi_0 = k_0 \xi_3 /k_3$ consistent with the paraxial approximation $\xi_3 \simeq \xi_r$, inserting this into Eqs. (\ref{eq: gouy_phase_lp}) and (\ref{eq: complex_beam_parameter}) yields  the traditional paraxial beam formula
\begin{eqnarray} 
\label{eq: gouy_phase_lp_spatial}
g_{lp} & = & (1+|l|+2p)\arctan \left( \xi_3/\xi_R \right),
\\
|w| & = & w_0\sqrt{1+ \left( \xi_3/\xi_R \right)^2} \, ,
\end{eqnarray}
The beam radius $|w|$ determines an important relation
\begin{equation} \label{eq: divergence_angle}
	\sin \theta_D \,\, = \,\, \lim\nolimits_{\xi_3 \rightarrow \infty}\left(|w|/\xi_3\right) 
	\,\, = \,\, 2 \, /\left(w_0k_3\right) 
\end{equation}
between the angular divergence of the beam, $\theta_D$, the longitudinal component of the wave vector and the beam radius at the waist. Figure 1 shows a comparison of curved non-paraxial to planar paraxial Gouy phase fronts calculated using Eqs. (\ref{eq: gouy_phase}) and (\ref{eq: gouy_phase_lp_spatial}) respectively. 
\begin{figure}[htp]
	\centering
	\includegraphics[width=8.2 cm]{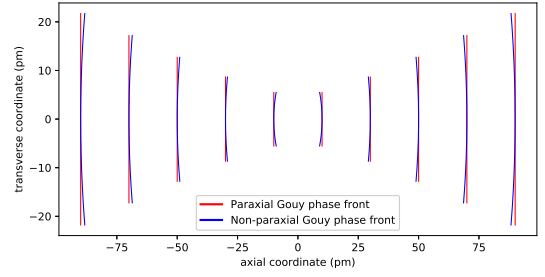}
	\caption{A comparison of paraxial (flat) and beyond paraxial (curved) Gouy phase fronts for a 100KeV electron beam. The
	beam radius of 5pm is set smaller than occurs in practice to accentuate the difference. }
	\label{fig:1}
\end{figure}

In order to estimate the magnitude of terms in the solution~(\ref{eq:
bispinor_explicit}), we evaluate the averaged probability density:
\begin{equation}
\nonumber
\langle |\Psi_\pm|^2 \rangle \,\, =  \,\, (b^2 + \hbar k_3^2) \langle |\Psi_{00}|^2
\rangle + 2\hbar^2 \kappa^2 \langle |\Psi_{01}|^2 \rangle + \frac{2\hbar^2}{w_0^2}\langle |\Psi_{10}|^2 \rangle \, ,
\end{equation}
where we use $\langle |\Psi_{lp}|^2 \rangle = C^2$ to confirm
$\langle j_0 \rangle = \langle |\Psi_\pm|^2 \rangle = 1$.
The cross terms vanish here since the products of the bi-spinors are identically zero. It now follows that
\begin{equation}
2\hbar^2 \kappa^2 \langle |\Psi_{01}|^2 \rangle
/\langle |\Psi_\pm|^2 \rangle \,\, = \,\,
2 \hbar^2 \kappa^2 C^2 \,\, < \,\, 10^{-8}
\end{equation}
owing to current imperfections in magnetic lenses that limit $w_0$
to values of about $50\;\mathrm{pm}$ or greater. For our further purposes,
it is therefore reasonable to drop the negligible term
in Eq. (\ref{eq: bispinor_explicit}) giving $\Psi_\pm(x_\mu, \xi_\mu)$ to be
\begin{equation} \label{eq: bispinor_useful}
\Psi_\pm \, = \, \left( \begin{array}{c}
b\chi_{\pm} \\
\pm \hbar k_3 \chi_{\pm}
\end{array} \right) \Psi_{00}
\pm \left( \begin{array}{c}
0 \\
\frac{\sqrt{2} \hbar}{w_0} \chi_{\mp}
\end{array} \right) \Psi_{10}
\end{equation}
and $C \simeq \sqrt{1 / (2\hbar k_0 b) }$ holding to a very high degree of accuracy. Equation
(\ref{eq: bispinor_useful}) completes the solution of the DE for the relativistic BHG beam of electrons
and enables evaluation of linear and angular momenta in the beam.

{\textit{Momentum and energy of the beam.---}
Some of relativistic beam solutions, such as Bessel \cite{Bliokh:11} and Volkov-Bessel modes \cite{Hayrapetyan-Karlovets}, although reasonable in other respects, actually
carry an infinite beam energy. BHG solutions, similar to LG modes, do not share this `problem'. In particular, the expectation values for the $4$-momentum in a beam are determined to be 
$p_\mu = \langle \Psi_{\pm}^{\dagger} \hat{p}_{\mu} \Psi_{\pm} \rangle = \hbar k_\mu$.
Inserting this result into the dispersion relation $k_\mu^\prime k^{\mu \prime} = m^2 c^2/\hbar^2$ we
obtain the averaged total energy $E = p_0 c$ of a single Dirac particle in a Gaussian beam
\begin{equation} \label{eq: expected_total_energy}
E \,\,
= \,\, +c\sqrt{p_\rho^2 + p_3^2 + m^2c^2} \, ,
\end{equation}
where $p_\rho^2 = \langle \Psi_{\pm}^{\dagger} \hat{p}_\rho^{2} \Psi_{\pm}\rangle =  2\hbar^2/w_0^2$ and $p_3^2 =\hbar^2 k_3^2$ denote the square values of radial and axial momentum respectively.
Equation (\ref{eq: expected_total_energy}) has been obtained elsewhere 
for a Klein-Gordon particle in a Gaussian beam, see Ref. \cite{RD2}, 
which also connects the
stored kinetic energy in the beam to the Bohm potential \cite{PRH1}.

\textit{Fractional angular momenta and non-trivial phase structure of the beam.---}
Expected values for angular momenta of an 
electron parallel to the beam axis can be calculated if the explicit forms of the spin angular momentum (SAM), 
$\hat{S}_{3} = \left(\hbar/2 \right) \textrm{diag} \left( \sigma_3, \sigma_3 \right)$, and OAM operators,
$\hat{L}_{3} = \left(\hbar/\imath \right) 
\left( \xi_1 \partial/\partial x_2 - \xi_2 \partial/\partial x_1\right)$, are employed. Direct evaluation of corresponding integrals leads to 
\begin{equation} \label{eq: expected_am}
\langle \Psi_{\pm}^{\dagger} \hat{S}_3 \Psi_{\pm} \rangle \,\, = \,\,   \left( 1 - \Delta \right)s\hbar \, , \quad \langle \Psi_{\pm}^{\dagger} \hat{L}_3 \Psi_{\pm} \rangle \,\, = \,\, \Delta s \hbar \, ,
\end{equation}
where $s=\pm \frac{1}{2}$, while
\begin{equation} \label{eq: expected_fam}
	\Delta \,\, \equiv \,\, \Delta \left( \theta_D \right) \,\, = \,\,
	\left( 1 - m c^2/E \right) \sin^2 \theta_D
\end{equation}
represents the intrinsic SOI term. There is a little need for us to dwell on these expressions 
since they look like a special case ($\ell = 0$) of more general relations \textit{but} derived from relativistic Bessel-beam solutions to the DE~\cite{Bliokh:11} (c.f.,~\cite{footnote2}).

Nonetheless, there are two subtle differences in the SOI terms for Bessel and BHG beams. (i) The Bessel-type solutions naturally contain non-paraxiality as a key feature, which is quantified by means of a parameter called opening angle $\theta_0 = \arcsin \left(\sqrt{k_1^2+k_2^2}/k\right)$ similarly appearing in the sine function \cite{Bliokh:11}. In our case of Eq. (\ref{eq: expected_fam}), we have the divergence angle instead, that carries \textit{more} information about the beam radius (at the waist) via Eq.~(\ref{eq: divergence_angle}). (ii) It is true that in the non-relativistic regime ($k \rightarrow 0$) SOI terms vanish for both types of beams. Due to the transverse localization of Gaussian modes, moreover, the first term of the SOI parameter depends \textit{explicitly} on the beam radius at the waist by virtue of Eq. (\ref{eq: expected_total_energy}). Combining these yields an explicit connection between the SOI term and the beamwidth, which can be simplified to 
\begin{eqnarray}
	\Delta & \approx & 2 \hbar^2/\left(m^2 c^2 w_0^2\right)
		\left[1 + 2/ \left( w_0^2 k_3^2 \right) \right]
\end{eqnarray}
for existing experimental setups, i.e., when $\hbar / \left(m c w_0\right) \ll 1$ for $w_0 > 50$ pm.
As seen, the SOI term vanishes for large beam radii \textit{independent} of the longitudinal momentum $\hbar k_3$, or else, 
for very small divergence angles (c.f., Eqs.~\ref{eq: divergence_angle} and \ref{eq: expected_fam}).

Another implication of expected angular momenta is that the focusing of relativistic Gaussian modes with bi-spinor structure will cause a fraction of angular momentum, $\Delta s \hbar$, to convert from the expected SAM to OAM and vice versa (see the left panel of Fig.~\ref{fig:2} for $s=1/2$). At the same time, the total angular momentum (TAM) of the beam, $\langle \Psi_{\pm}^{\dagger} \hat{J}_3 \Psi_{\pm} \rangle = s \hbar$ with $\hat{J}_3 = \hat{L}_3 + \hat{S}_3$, is conserved along the propagation direction as depicted by the straight green dash-dotted line. As the divergence angle increases from $0$ to $\pi/2$, the stake $\Delta s \hbar$ starts disappearing from the SAM and reappearing as fOAM resulting in the \textit{fractional spin-to-orbit conversion}. For $\theta_D = \pi / 2$, the fractional SAM (fSAM) and fOAM parts contribute in the conserved TAM with the shares $m c^2 s \hbar / E$ and $\left( 1 - m c^2 / E \right)s \hbar$, respectively. Depending on the electron kinetic energy, the spin-to-orbit conversion occurs either fully (for $0.5\;\mathrm{MeV}$), so that the shares by angular momenta are equal, or partially (e.g., for $0.1\;\mathrm{MeV}$) due to the gradual decrease of the SOI term.

\begin{figure}[htp]
	\centering
	\includegraphics[width=4.2 cm]{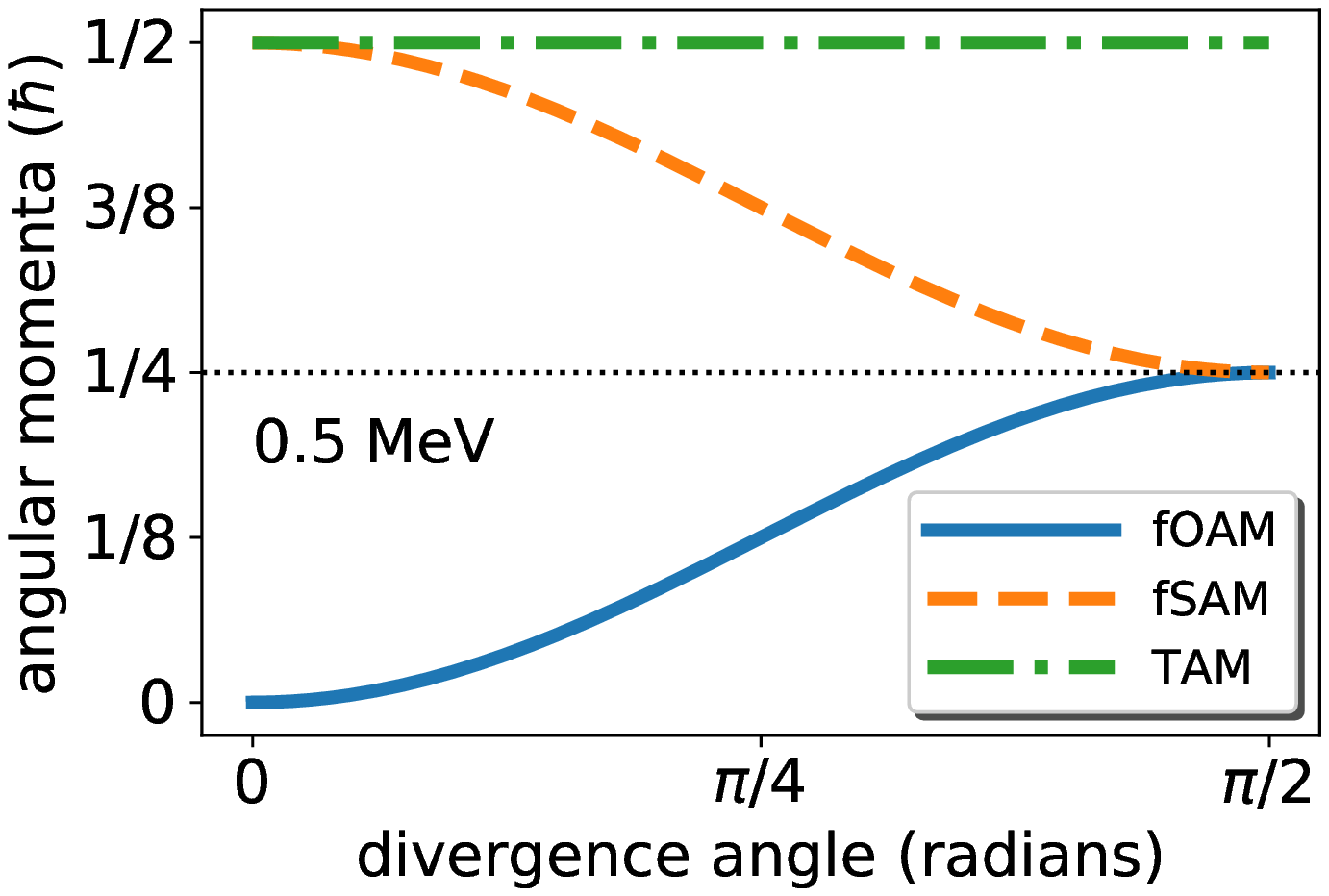}
	\includegraphics[width=4.2 cm]{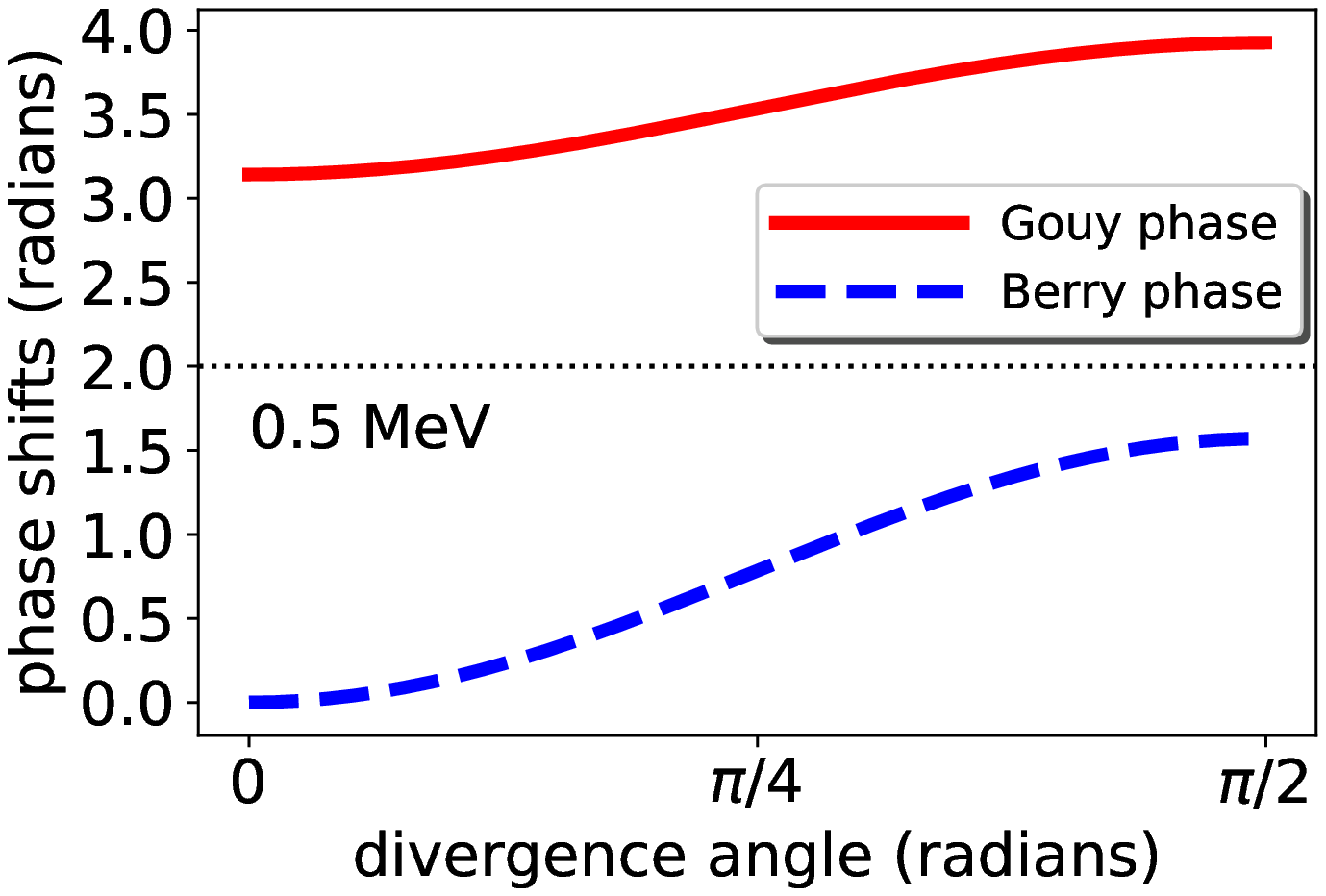}
	\includegraphics[width=4.2 cm]{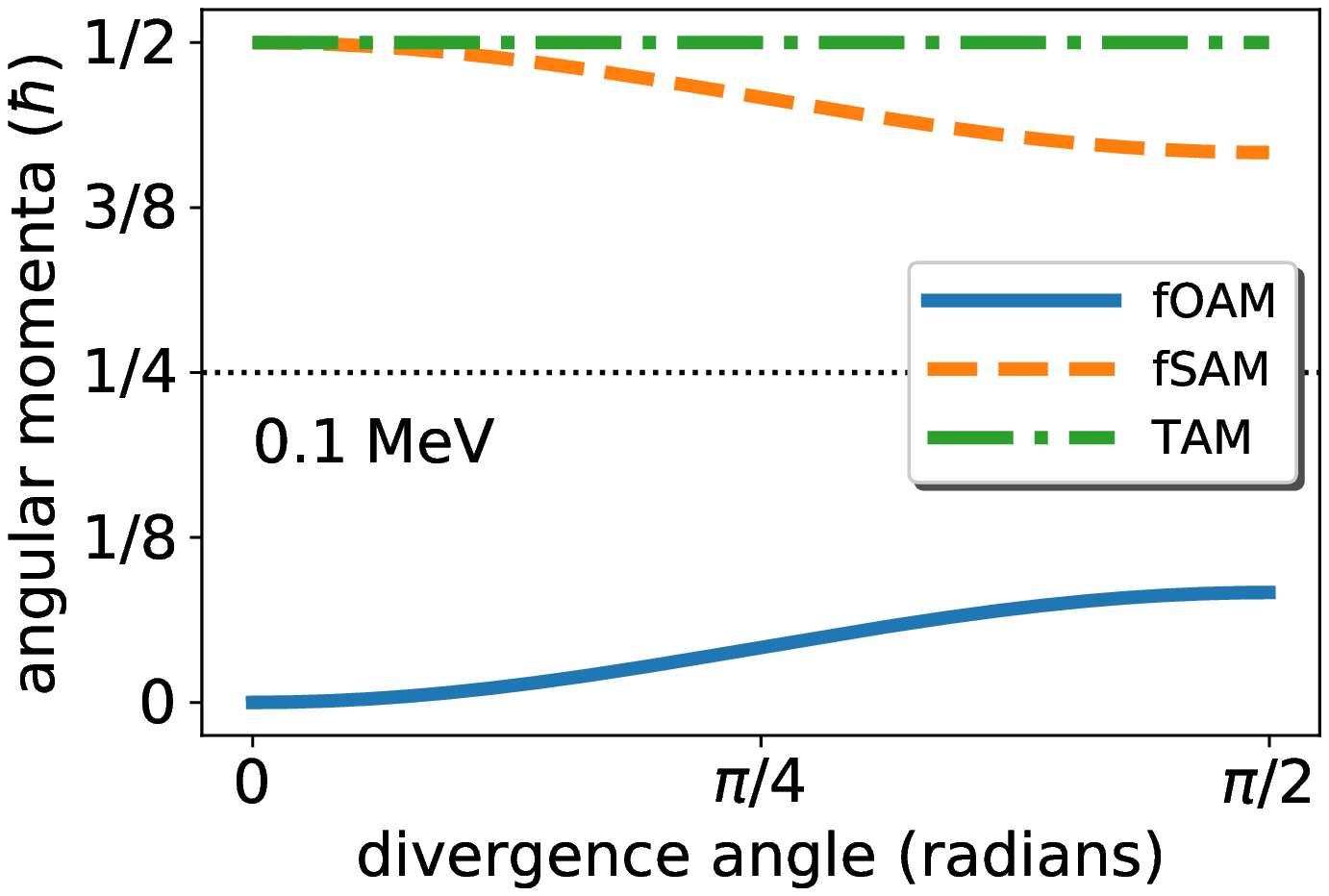}
	\includegraphics[width=4.2 cm]{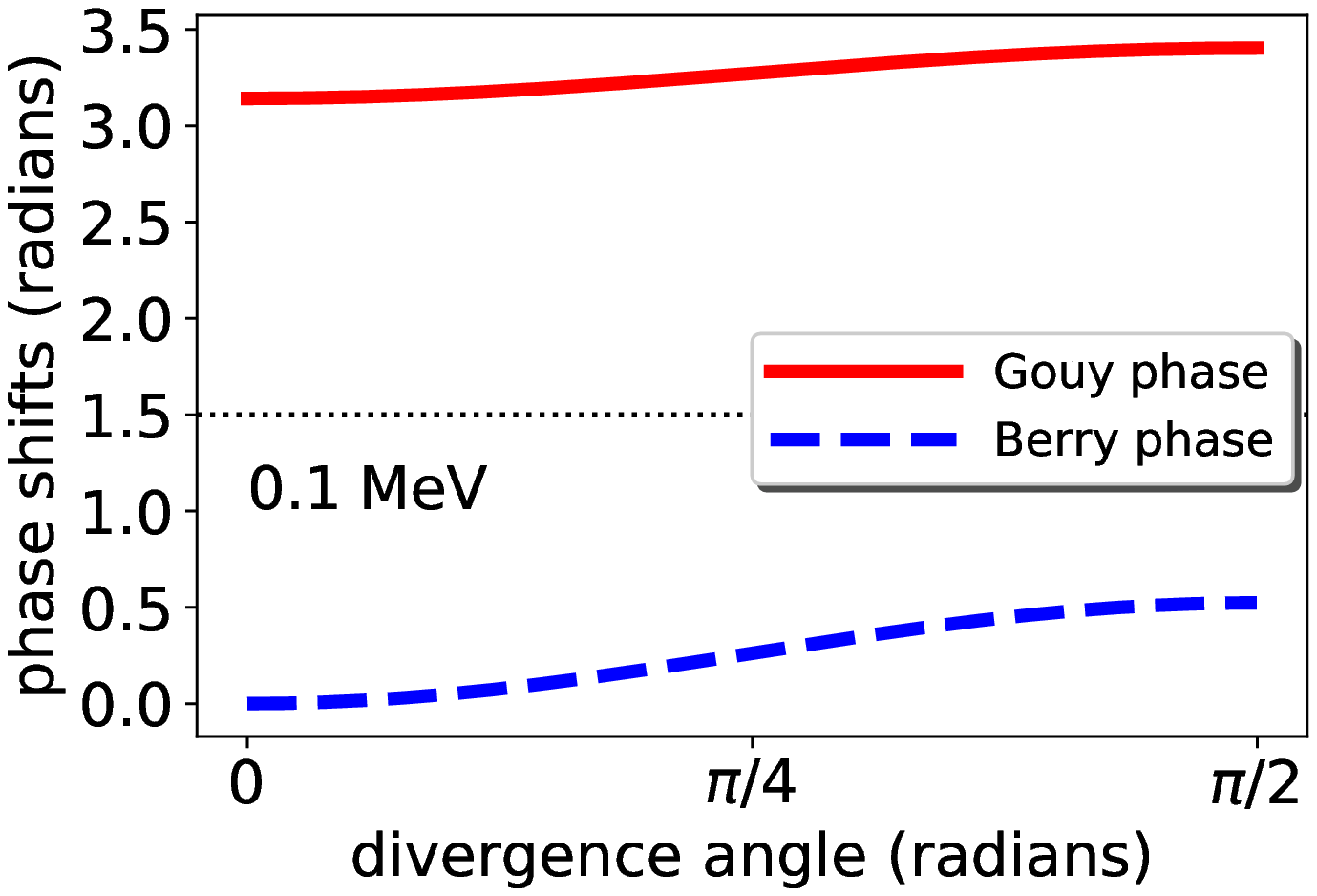}
	\caption{Fractional angular momenta (left panel), the total Gouy  
		phase shift and Berry phase (right panel) as functions
		of the divergence angle
		for spin-up electrons with
		a kinetic energy of $0.5\;\mathrm{MeV}$ (upper panel) and 
		of $0.1\;\mathrm{MeV}$ (lower panel).}
	\label{fig:2}
\end{figure}

Variations in the beamwidth over the infinite length of the beam axis represent an adiabatic cycle in the sense that the beam divergence beyond the waist undoes the convergence occurring ahead of the waist. Such an adiabatic cycle, being inherently characterized by the divergence angle, contributes in the SOI due to the tight focusing of the BHG beam by means of non-zero higher mode bi-spinors proportional to $\Psi_{01}$ and $\Psi_{10}$ (see Eq.~(\ref{eq: bispinor_explicit})). Over the course of such a cyclic adiabatic process, the beam accumulates also a Berry (geometric) phase from these higher mode bi-spinors. The Berry phase can be evaluated exactly in the same manner as for relativistic Bessel beams by making use of the so-called Foldy-Wouthuysen momentum representation \cite{Bliokh:11}. Following Bliokh {\it et al}. we may therefore write the Berry phase as gained due to the non-trivial fOAM as $\gamma_B = 2 \pi \Delta s$.

For a tightly focused BHG beam, the expected Gouy phase 
\begin{equation}
\label{Gouy_phase_shift} 
	\bar{g}_{T} \,\, = \,\, 
	\sum\nolimits_{lp} \langle \Psi_{\pm}^{\dagger} g_{lp} \Psi_{\pm} \rangle
	\,\, = \,\,
	\left[1+ 0.5 \, \Delta(\theta_D)\right] g_{00}
\end{equation}
is larger than would be the case for a pure Gaussian beam owing to the SOI. The total Gouy phase shift from far field to far field in the beam is therefore given by
\begin{equation}
	\mu_T \,\, =\,\, 
	\lim_{\xi_3 \rightarrow \infty} (\bar{g}_{T}) - \lim_{\xi_3 \rightarrow -\infty}(\bar{g}_{T}) 
	\,\, = \,\, \pi + 0.5 \, |\gamma_B| \, ,
\end{equation}
showing that the Gouy phase as well as the fOAM increase in direct proportion to the Berry phase, while the fSAM decreases. The right panel of Fig.~\ref{fig:1} illustrates the growth of both 
phases dependent on the divergence angle and the electron kinetic energy. The Berry phase rises as the SOI term increases and retains its maximum value along with the Gouy phase at $\theta_D = \pi / 2$.
The presence of non-vanishing expectation values for the transverse components of the linear momentum implies that the relevant phase angle to consider in this case is not the azimuth phase associated with the fOAM but the Gouy phase~\cite{feng2001-yang2006}. 
It thus appears from our results that the Berry phase for this cycle is not the baseline Gouy phase for the paraxial beam but the fractional increase in the Gouy phase above the baseline value that can be explicitly attributed to the adiabatic cycle. Our evidence for this assertion being the direct proportionality between the Gouy phase and the fOAM value is shown in Eq. (\ref{Gouy_phase_shift}).

{\textit{Discussion}.---} Dirac published his quantum theory of the electron in 1928 \cite{PAMD}. Some ninety years later, it is now being extensively applied to understand the effects of transverse localization on electron beams. A significant progress has been made in this direction by deriving (either exact or approximated) solutions to the DE for Bessel \cite{Bliokh:11,Hayrapetyan-Karlovets} and LG beams \cite{Karlovets:18,BB,SMB}. In our paper, we derive a new and exact BHG solution to the DE, which possesses the full relativistic nature of the beam propagation as it takes into account the $4$-position of the beam waist~\cite{footnote3}. This has enabled us to calculate the energy-momentum, fractional angular momenta and the Gouy phase in the beam and demonstrate the presence of the intrinsic SOI leading to the fractional spin-to-orbit conversion. Remarkably, both the fOAM and the Gouy phase are directly proportional to the Berry phase. This both corroborates the earlier finding of Bliokh {\it et al.} \cite{Bliokh:11} for fOAM and takes it a step further with our inclusion of Gouy phase into the evolving understanding of the role geometric phase has to play in relativistic electron beams, predicted earlier for a non-relativistic Gaussian beam~\cite{SM} and demonstrated recently for optical rays~\cite{Malhotra:18}. Additionally, we have found that Gouy phase fronts that have traditionally thought to be planar are in fact curved. This curvature is most apparent in the far fields of strongly diffracting beams.

\begin{acknowledgements}
I. G. da Paz thanks Grant No. 307942/2019-8 from CNPq.
\end{acknowledgements}


\begin{thebibliography}{99}

\bibitem{Bliokh:11}
K. Y. Bliokh, M. R. Dennis, and F. Nori, {Phys. Rev. Lett.}
\textbf{107}, 174802 (2011).

\bibitem{Hayrapetyan-Karlovets}
A. G. Hayrapetyan, O. Matula, A. Aiello, A. Surzhykov, and S.
Fritzsche, {Phys. Rev. Lett.} \textbf{112}, 134801 (2014);
D. V. Karlovets, {Phys. Rev. A} \textbf{86}, 062102 (2012).

\bibitem{Bliokh:07}
K. Y. Bliokh, Y. P. Bliokh, S. Savel'ev, and F. Nori,
{Phys. Rev. Lett.} \textbf{99}, 190404 (2007);

\bibitem{Schattschneider:11}
P. Schattschneider and J. Verbeeck, {Ultramicroscopy} \textbf{111},
1461 (2011).

\bibitem{Bliokh:17}
K. Y. Bliokh, M. R. Dennis, and F. Nori,
{Phys. Rev. A} \textbf{96}, 023622 (2017).

\bibitem{Karlovets:18}
D. Karlovets, {Phys. Rev. A} \textbf{98}, 012137 (2018). 

\bibitem{BB}
I. Bialynicki-Birula and Z. Bialynicka-Birula, {Phys. Rev.
	Lett.} \textbf{118}, 114801 (2017).

\bibitem{SMB}
S. M. Barnett, {Phys. Rev. Lett.} \textbf{118}, 114802 (2017).

\bibitem{BB-comment}
I. Bialynicki-Birula and Z. Bialynicki-Birula,
{Phys. Rev. Lett.} \textbf{119}, 029501 (2017).

\bibitem{SR} M. Saleem and M. Rafique, \emph{Special Relativity Applications to Particle Physics and the Classical Theory of Fields} (Ellis Horwood,
1992).

\bibitem{AK} A. Komar, {Phys. Rev. D} \textbf{18}, 1887 (1978).

\bibitem{CA} H. W. Crater and P. Van Alstine, {Phys. Rev. D} \textbf{36},3007 (1987).


\bibitem{RD1}
R. Ducharme and I. G. da Paz, {Phys. Rev. A} \textbf{92}, 023853
(2015).

\bibitem{RD2}
R. Ducharme and I. G. da Paz, {Phys. Rev. A} \textbf{94}, 023822
(2016).

\bibitem{RD3} R. Ducharme, {Prog Electromagn Res M} \textbf{42}, 39
(2015).

\bibitem{RE}
R. Erni, M. D. Rossell, C. Kisielowski, and U. Dahmen, {Phys. Rev.
	Lett}. \textbf{102}, 096101 (2009).

\bibitem{EVB-exp}
M. Uchida and A. Tonomouro, {Nature} (London) \textbf{464}, 737
(2010); J. Verbeeck, H. Tian, and P. Schattschneider, {Nature}
(London) \textbf{467}, 301 (2010); B. J. McMorran, A. Agrawal, I. M.
Anderson, A. A. Herzing, H. J. Lezec, J. J. McClelland, and J.
Unguris, {Science} \textbf{331}, 192 (2011); J. Verbeeck, P.
Schattschneider, S. Lazar, M. Stoger-Pöllach, S. Löffler, A.
Steiger-Thirsfeld, and G. Van Tendeloo, {Appl. Phys. Lett.}
\textbf{99}, 203109 (2011);
V. Grillo, E. Karimi, G. C. Gazzadi, S. Frabboni, M. R. Dennis, and
R. W. Boyd, {Phys. Rev. X} \textbf{4}, 011013 (2014);
E. Mafakheri, A. H. Tavabi, P.-H. Lu, R. Balboni, F. Venturi, C. Menozzi, G. C. Gazzadi, S. Frabboni, A. Sit, R. E. Dunin-Borkowski, E. Karimi, and V. Grillo, {Appl. Phys. Lett.} \textbf{110}, 093113 (2017);
G. M. Vanacore, G. Berruto, I. Madan, E. Pomarico, P. Biagioni, R. J. Lamb, D. McGrouther, O. Reinhardt, I. Kaminer, B. Barwick, H. Larocque, V. Grillo, E. Karimi, F. J. Garc\'ia de Abajo, and F. Carbone, {Nat. Materials} 
\textbf{18}, 573 (2019).

\bibitem{review-EVB}
K. Y. Bliokh, I. P. Ivanov, G. Guzzinati, L. Clark, R. Van Boxem,
A. B\'ech\'e, R.Juchtmans, M. A. Alonso, P. Schattschneider, F. Nori, and
J. Verbeeck, {Phys. Rep.} \textbf{690}, 1 (2017);
J. Harris, V. Grillo, E. Mafakheri, G. C. Gazzadi, S. Frabboni, R. W. Boyd,
and E. Karimi, {Nat. Phys.} \textbf{11}, 629 (2015);
S. M. Lloyd, M. Babiker, G. Thirunavukkarasu, and J. Yuan
{Rev. Mod. Phys.} \textbf{89}, 035004 (2017).

\bibitem{EVB-in-external-fields}
K. Y. Bliokh, P. Schattschneider, J. Verbeeck, and F. Nori,
{Phys. Rev. X} \textbf{2}, 041011 (2012);
C. Greenshields, R. L. Stamps, and S. Franke-Arnold,
{New J. Phys.} \textbf{14}, 103040 (2012);
G. M. Gallatin and B. McMorran, {Phys. Rev. A} \textbf{86}, 012701 (2012);
M. Babiker, J. Yuan, and V. E. Lembessis, {Phys. Rev. A} \textbf{91}, 013806 (2015);
K. van Kruining, A. G. Hayrapetyan, and J. B. G\"otte,
{Phys. Rev. Lett.} \textbf{119}, 030401 (2017);
A. J. Silenko, P. Zhang, and L. Zou, {Phys. Rev. Lett.}
\textbf{119}, 243903 (2017) \textit{ibid.} \textbf{121} 043202 (2018).

\bibitem{EVB-with-scattering}
I. P. Ivanov, {Phys. Rev. D.} \textbf{83}, 093001 (2011) \textit{ibid.} \textbf{85}, 076001 (2012);
V. Serbo, I. P. Ivanov, S. Fritzsche, D. Seipt, and A. Surzhykov,
{Phys. Rev. A} \textbf{92}, 012705 (2015);
I. P. Ivanov, D. Seipt, A. Surzhykov, and S. Fritzsche,
{Phys. Rev. D} \textbf{94}, 076001 (2016);
I. P. Ivanov, N. Korchagin, A. Pimikov, and P. Zhang, {Phys. Rev. Lett.}. \textbf{124}, 192001 (2020).

\bibitem{EVB-with-radiation}
O. Matula, A. G. Hayrapetyan, V. G. Serbo, A. Surzhykov, and S. Fritzsche, 
{New J. Phys} \textbf{16}, 053024 (2014);
V. A. Zaytsev, V. G. Serbo, and V. M. Shabaev, {Phys. Rev. A} \textbf{95}, 012702 (2017).

\bibitem{manipulation-via-twist}
D. G. Grier, {Nature}, \textbf{424}, 810 (2003); 
E. G. Abramochkin, S. P. Kotova, A. V. Korobtsov, N. N. Losevsky, 
A. M. Mayorova, M. A. Rakhmatulin, and V. G. Volostnikov,
{Las. Physics} \textbf{16}, 842 (2006);
K. Toyoda, K. Miyamoto, N. Aoki, R. Morita, and T. Omatsu,
{Nano Lett.}, \textbf{12}, 3645 (2012);
W. Brullot, M. K. Vanbel, T. Swusten, and Thierry Verbiest,
{Sci. Adv.} \textbf{2}, e1501349 (2016).

\bibitem{review-twisted-light}
L. Allen, M. W. Beijersbergen, R.J.C. Spreeuw, and
J.P. Woerdman, {Phys. Rev. A} \textbf{45}, 8185 (1992); L. Allen, M.
Padgett, and M. Babiker, {Prog. Opt.} \textbf{39 }, 291 (1999);
G. Molina-Terriza, J.P. Torres, and L. Torner, 
{Nat. Phys.} \textbf{3}, 305 (2007);
S. Franke-Arnold, L. Allen, and M.J. Padgett, 
{Laser \& Photon. Rev.} \textbf{2}, 299 (2008);
K. Y. Bliokh and A. Aiello, {J. Opt.} \textbf{15}, 014001 (2013);
D. L. Andrews and M. Babiker, \textit{Angular Momentum of Light}, 
Cambridge University Press 2013;
M. J. Padgett, {Opt. Express} \textbf{25}, 11265 (2017).

\bibitem{twisted-light-other}
K. Y. Bliokh, M. A. Alonso, E. A. Ostrovskaya, and A. Aiello,
{Phys. Rev. A} \textbf{82}, 063825 (2010);
L. Marrucci, E. Karimi, S. Slussarenko, B. Piccirillo, E. Santamato, E.
Nagali, and F. Sciarrino, {J. Opt.}, \textbf{13}, 064001 (2011);
O. Matula, A. G. Hayrapetyan, V. G. Serbo, A. Surzhykov, and S. Fritzsche,
{J. Phys. B} \textbf{46}, 205002 (2013);
H. M. Scholz-Marggraf, S. Fritzsche, V. G. Serbo, A. Afanasev, and A. Surzhykov
{Phys. Rev. A} \textbf{90}, 013425 (2014);
A. Aiello, P. Banzer, M. Neugebauer, and G. Leuchs,
{Nat. Photon.} \textbf{9}, 789 (2015);
M. Krenn, J. Handsteiner, M. Fink, R. Fickler, R. Ursin, M. Malika,
and A. Zeilinger, {Proc. Natl. Acad. Sci. U.S.A.}
\textbf{113}, 13648 (2016);
M. Erhard, R. Fickler, M. Krenn, and A. Zeilinger,
{Light: Sci. \& Applications} \textbf{7}, 17146 (2018).

\bibitem{FOAM} 
M. V. Berry, {J. Opt. A} \textbf{6}, 259 (2004).

\bibitem{FOAM-exp}
J. B. Götte, K. O. Holleran, D. Preece, F. Flossmann, S. Franke-Arnold, S. Barnett, and M. Padgett, {Opt. Express}  \textbf{16}, 993 (2008);
D. P. O'Dwyer, C. F. Phelan, Y. P. Rakovich, P. R. Eastham, J. G. Lunney, and J. F. Donegan, {Opt. Express}, \textbf{18}, 16480 (2010).

\bibitem{gouy1} 
L. G. Gouy, {C. R. Acad. Sci. Paris} \textbf{110}, 1251 (1890);
L. G. Gouy, {Ann. Chim. Phys. Ser. 6} \textbf{24}, 145 (1891).

\bibitem{gouy2}
T. D. Visser and E. Wolf, {Opt. Commun.} \textbf{283}, 3371 (2010);
J. Yang and H. G. Winful, {Opt. Lett.} \textbf{31}, 104 (2006); 
R. W. Boyd, {J. Opt. Soc. Am.} \textbf{70}, 877 (1980); 
P. Hariharan and P. A. Robinson, {J. Mod. Opt.} \textbf{43}, 219 (1996); 
S. Feng, H. G. Winful, and R. W. Hellwarth, {Opt. Lett.} \textbf{23}, 385
(1998); 
D. Chauvat, O. Emile, M. Brunel, and A. Le Floch, {Am. J.
Phys.} \textbf{71}, 1196 (2003); 
I. G. da Paz, P. L. Saldanha, M. C. Nemes, and J. G. Peixoto de Faria, 
{New J. Phys.} \textbf{13}, 125005 (2011);
X. Pang, D. G. Fischer, and T. D.
Visser, {Opt. Lett.} \textbf{39}, 88 (2014).

\bibitem{feng2001-yang2006} 
S. Feng and H. G. Winful, {Opt. Lett.} \textbf{26}, 485 (2001);
J. Yang and H. G. Winful, {Opt. Lett.} \textbf{31}, 104 (2006).

\bibitem{BA-PH} 
H. Bateman, {Proc. London Math. Soc.} \textbf{7}, 77 (1909)
\textit{ibid.} \textbf{8}, 223 (1910) \textit{ibid.} \textbf{8}, 469 (1910); 
P. Hillion, {J. Math. Phys.} \textbf{33}, 2749 (1992).

\bibitem{BS-APK} 
I. M. Besieris and A. M. Shaarawi, {J. Electromagn. Waves and Appl.} \textbf{16}, 1047 (2002)
\textit{ibid.} {Opt. Express} \textbf{27}, 792 (2019);
A. P. Kiselev, {J. Math. Phys.} \textbf{41}, 1934 (2000)
\textit{ibid.} {Optics and Spectroscopy.} \textbf{102}, 603 (2007).

\bibitem{PAMD} P.A.M Dirac, {Proc. Roy. Soc.} \textbf{778}, 223 (1928).

\bibitem{footnote1}
Throughout the text, 
$\langle ... \rangle \equiv 
 \iint_{-\infty}^{+\infty} ... d\xi_1 d\xi_2$. 
As $j_\mu\left(\xi_1 , \xi_2 , \xi_3 + \xi_0 \right)$ is a conserved
quantity, we expect and find that $\langle j_\mu \rangle$ is independent
of $\xi_3$ and $\xi_0$ even though no averaging over these
coordinates is carried out.

\bibitem{PRH1} P. R. Holland, {Found. Phys} \textbf{45}, 134 (2015).

\bibitem{footnote2}
Similar SOI term (proportional to the absolute value of OAM) also appears elsewhere as a contribution to the spin part of the magnetic moment of some suitably constructed relativistic vortex wave packet [c.f. Eqs. (23) and (47) of Ref. \cite{Karlovets:18}].

\bibitem{footnote3} Such an approach could be developed also for other types of matter vortex beams~\cite{other_VB}.

\bibitem{other_VB}
A. G. Hayrapetyan, O. Matula, A. Surzhykov, and S. Fritzsche,
{Eur. Phys. J. D} \textbf{67}, 167 (2013);
A. G. Hayrapetyan and S. Fritzsche, {Phys. Scr.}
\textbf{T156}, 014067 (2013);
C. W. Clark, R. Barankov, M. G. Huber, M. Arif, D. G. Cory, and D. A. Pushin,
{Nature} \textbf{525}, 504 (2015);
K. Y. Bliokh and F. Nori, {Phys. Rev. B} \textbf{99}, 174310 (2019);
I. Rond\'on and D. Leykam, {J.\ Phys.: Condens. Matter} \textbf{32}, 104001 (2020).

\bibitem{SM} R. Simon and N. Mukunda, {Phys. Rev. Lett.} \textbf{70}, 880
(1993).

\bibitem{Malhotra:18}
T. Malhotra, R. Guti\'errez-Cuevas, J. Hassett, M. R. Dennis, A. N. Vamivakas, and M. A. Alonso, {Phys. Rev. Lett.} \textbf{120}, 233602 (2018).

\end{thebibliography}
\end{document}